\documentclass[twocolumn, prB]{revtex4-1}
\usepackage{graphicx}
\usepackage[
	pdftitle={Cyanographone and Isocyanographone -- two asymmetrically functionalized graphene pseudohalides and their potential use in chemical sensing},
	pdfsubject={Article},
	pdfauthor={Lukas Eugen Marsoner Steinkasserer, Vincent Pohl, and Beate Paulus},
	pdfdisplaydoctitle=true,
	colorlinks=true,allcolors = blue
]{hyperref}
\pdfoutput=1

\begin{document}
\title{Cyanographone and Isocyanographone -- two asymmetrically functionalized graphene pseudohalides and their potential use in chemical sensing}

\author{Lukas Eugen \surname{Marsoner Steinkasserer}}
\email{marsoner@zedat.fu-berlin.de}
\affiliation{Institut f\"{u}r Chemie und Biochemie, Freie Universit\"{a}t Berlin, Takustra\ss e 3, D-14195 Berlin, Germany}

\author{Vincent Pohl}
\affiliation{Institut f\"{u}r Chemie und Biochemie, Freie Universit\"{a}t Berlin, Takustra\ss e 3, D-14195 Berlin, Germany}

\author{Beate Paulus}
\affiliation{Institut f\"{u}r Chemie und Biochemie, Freie Universit\"{a}t Berlin, Takustra\ss e 3, D-14195 Berlin, Germany}

\date{\today}
\maketitle

\section{Abstract}
Graphene pseudohalides are natural candidates for use in molecular sensing due to their greater chemical activity as compared to both graphene halides and pristine graphene. Though their study is still in its infancy, being hindered until recently by the unavailability of both selective and efficient procedures for their synthesis, they promise to considerably widen the application potential of chemically modified graphenes. Herein, we employ vdW-DFT to study the structural and electronic properties of two selected graphene pseudohalides namely cyanographone and isocyanographone and investigate the potential use of the latter as a chemical sensor via electron transport calculations.

\section{Introduction}
In the wake of the so aptly termed \emph{rise of graphene} \cite{geim2007rise, Novoselov2004}, recent years have seen a growing interest in modifying graphene, overcoming its weaknesses, and tailoring its properties to specific applications. A promising avenue towards that goal consist in its functionalization by appropriately chosen chemical groups \cite{Chua2013, Karlick2017, Liao2014, Eigler2014, Robinson2010, Criado2015, Kuila2012, Georgakilas2012, Ivanovskii2012, Johns2013, Karlick2013, Tuek2017, Bakandritsos2017, Pumera2017, Bourlinos2012, Setaro2017, Paulus2013, Liu2012}. Among these, atomic substituents such as H, F, Cl and Br have attracted large amounts of attention, especially in the theoretical community, as they allow for a relatively simple and controlled functionalization and lend themselves well to computational study \cite{Zhang2012,Pykal2016}.

Though the interest in graphene halides has been extensive, interest in graphene pseudohalides has begun to emerge only rather recently \cite{Chronopoulos2017, Pumera2017, Karlick2013}. This is in part due to the fact that, until lately, no synthetic protocols were in place which allowed for the efficient and selective synthesis of such derivatives \cite{Bakandritsos2017, Chronopoulos2017, Pumera2017}. Pseudohalogens, e.g. N$_3$, SCN, CN, are highly attractive as graphene additives as they offer much greater chemical activity than true halogens, while at the same time presenting many of the attractive chemical properties of halogen atoms. Though there is in principle a wide variety of pseudohalides, herein we will focus solely on the cyano group (CN) and two possible derivatives that can be formed from it, i.e., cyanographene (GrCN) and isocyanographene (GrNC) in which the cyano group binds to the graphene layer via the carbon and nitrogen atom respectively.

Cyanographene has recently been synthesized for the first time by Bakandritsos and coworkers via the reaction of NaCN with fluorographene \cite{Bakandritsos2017}. While experimental results indicate a functionalization degree of $13 - 15$\%\, herein we will only consider stoichiometric derivatives. Moreover, though symmetrically functionalized materials are certainly interesting they, just as most graphene halides, have the disadvantage of being rather wide band gap semiconductors -- making them unappealing for many potential applications. One way to overcome this problem consists in asymmetrical functionalization on just one side of the graphene layer. We will, in analogy to half-hydrogenated graphene, use the suffix \emph{--one} to indicate such half-functionalized graphene pseudohalides. Half-cyanated graphene will therefore be termed \emph{cyanographone} while half-isocyanated graphene is referred to as \emph{isocyanographone}. Apart from analyzing the structural and electronic properties of the pristine and BN adsorbed cyano- and isocyanographone layers we will, for the case of cyanographone, see how adsorption of gas molecules can significantly alter its electron transport properties rendering it an interesting target for the development of nanoscale chemical sensors.

Gas sensors play a crucial role in a variety of fields, ranging from scientific research to industrial or domestic activities \cite{Capone2004, Sberveglieri1992, Moseley1997}. In recent years a growing number of researchers have become interested in exploiting the attractive properties of 2D materials such as their extremely high surface to volume ratios, for the detection of rarefied gases with the promise of achieving measurement accuracy down to individual molecules \cite{Varghese2015, Kou2014, Abbas2015, Lee2013}. 
One of the main challenges facing the application of 2D materials to gas sensing though, consist in selectively detecting specific types of molecules. While e.g. great progress has been made in the field of graphene-based sensors \cite{Schedin2007,Wang2015, Yuan2013, Huang2008, Robinson2008, Wang2015, Bogue2014, He2012}, most common gas species are only weakly adsorbed on the graphene surface \cite{Leenaerts2008, Wehling2008}, thereby increasing the risk of cross-sensitivity. 
Using electron transport calculations we will herein show how a hypothetical BN adsorbed cyanographone (GrCN@BN) based sensor is able to selectively detect CO and H$_2$S in the presence of a number of common atmospheric gases (N$_2$, CO$_2$, and O$_2$).

\section{Computational Details}
\label{sec:comp}

Calculations in this work were performed using the \textsc{GPAW} program \cite{Mortensen2005, Enkovaara2010, Marques2012, Blchl1994}. If not otherwise stated, all calculations further employed the libvdwxc \cite{Larsen2017} implementation of the vdW-DF-CX functional by Berland and Hyldgaard \cite{Berland2014a, Berland2014b, RomnPrez2009, Gharaee2017} which allows for a consistent description of non-local vdW-forces within the framework of density functional theory. We note at this point that, in the case of spin-polarized calculations, the total density was used to evaluate the non-local vdW correlation.

During structure optimization, atomic positions were relaxed until the remaining forces were below 0.05 eV/$\mathring{\mathrm{A}}$, while unit cell parameters were obtained from parabolic fits of energies around the minimum. Unless otherwise stated, all calculations were performed using a real-space grid representation of the wave functions with a grid spacing of 0.15 $\mathring{\mathrm{A}}$ (rounded slightly in order to match the unit cell) and employed at least a $24\times24$ $\Gamma$-centered Monkhorst-Pack grid.

Structure optimizations on supercells used to study molecular adsorption were done using a LCAO expansion of the wave functions \cite{Larsen2009} employing a double-$\zeta$ polarized (dzp) basis set. Unit cell parameters were kept constant during the relaxation while atomic positions were again allowed to relax until remaining forces were below 0.05 eV/$\mathring{\mathrm{A}}$. Binding energies were again obtained using a real-space grid representation of the wave functions. While structure optimizations employed a $\Gamma$-centered $4\times4$ Monkhorst-Pack grid for the $3\times3$ supercell, in the case of binding energies a $\Gamma$-centered $8\times8$ grid was used.

Nudged elastic band (NEB) calculations on transition paths and activation energies for molecular chemisorption reactions where performed using the \textsc{ASE} \cite{Bahn2002, LarsenASE} implementation of the method (see ref. \cite{PhysRevLett.72.1124, Jonsson385nudgedelastic} and \cite{LarsenASE} for details). Relaxations of the individual images employed the same parameters and methods as those used during relaxations of molecules on surfaces. After determination of the transition paths a final single point calculation was performed on all images using real-space grid representation of the wave functions in order to improve reaction energetics. This final calculation also employed a $\Gamma$-centered $4\times4$ Monkhorst-Pack grid.

Finally, electronic transport calculations were done via the \textsc{ASE} implementation of the nonequilibrium Green's function formalism (see ref. \cite{Strange2008, Thygesen2003, Thygesen2005a, Thygesen2005b} and \cite{LarsenASE} for details on the method and implementation). The Hamiltonians and overlap matrices were obtained from LCAO calculations employing the the vdW-DF-CX functional with a dzp basis set. The computational cell was sampled using 4 Monkhorst-Pack $\mathbf{k}$-points along the periodic direction.

\section{Results and Discussion}

\subsection{Symmetric and asymmetric functionalization}

\begin{figure}[h]
\centering
  \includegraphics[width=0.48\textwidth]{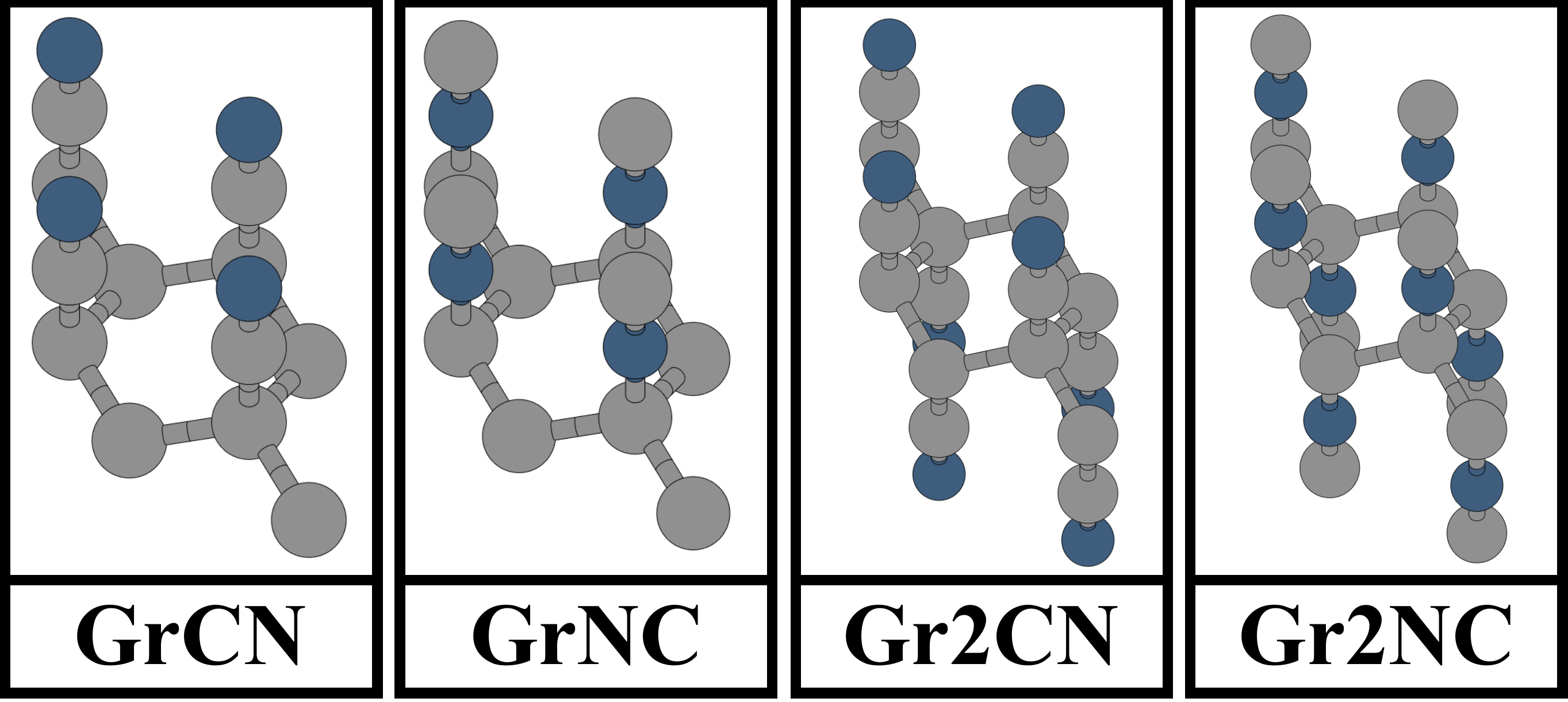}
  \caption{Summary of the different types of functionalization considered, together with the associated designations which will be used throughout this work. GrCN and GrNC refer to cyanographone and isocyanographone while labels Gr2CN and Gr2NC are used to indicate cyanographene and isocyanographene respectively. In all cases, carbon atoms are drawn in gray while nitrogen atoms are shown in dark blue.}
\label{systems}
\end{figure}

As shown in fig. \ref{systems}, we initially considered both symmetrically (Gr2CN, Gr2NC) and asymmetrically saturated (GrCN, GrNC) graphene derivatives. The former can be expected to display an electronic structure analogous to that of \emph{ordinary} graphene halides such as fluorographene \cite{Zboil2010, Nair2010, Robinson2010} and graphane \cite{Elias2009, Sofo2007, Sluiter2003}, characterized by rather large band gaps. To verify this, we calculated the quasiparticle gaps of the aforementioned systems using the {\mbox{GLLB-SC}} potential by Kuisma \emph{et al.} \cite{Kuisma2010, Gritsenko1995} which has been shown to yield excellent agreement with experimental results for a large number of systems \cite{Castelli2012}. 

As expected, we find Gr2CN/Gr2NC both to be wide band gap semiconductors with Gr2CN having a {\mbox{GLLB-SC$+\Delta_{\mathrm{xc}}$}} quasiparticle gap equal to 3.9 eV while Gr2NC shows a gap of 3.4 eV. Given that their large band gaps make them unappealing for a number of potential applications such as photovoltaics or chemical sensing, we will herein not consider the fully saturated cases any further, focusing rather on the more interesting half-functionalized systems. 
These asymmetrically saturated structures contain an unpaired electron on one of the two graphene sublattices (the other being functionalized by the cyano group), making them analogous to half-hydrogenated/-fluorinated graphene, i.e., graphone/fluorographone \cite{aolu2017, Buonocore2016, Zhou2011, Podlivaev2011, woellner2016graphone1, woellner2016graphone2, Peng2014, ahin2011, Hemmatiyan2014, Ray2014, Boukhvalov2010}. 

\begin{figure}[h]
\centering
  \includegraphics[width=0.48\textwidth]{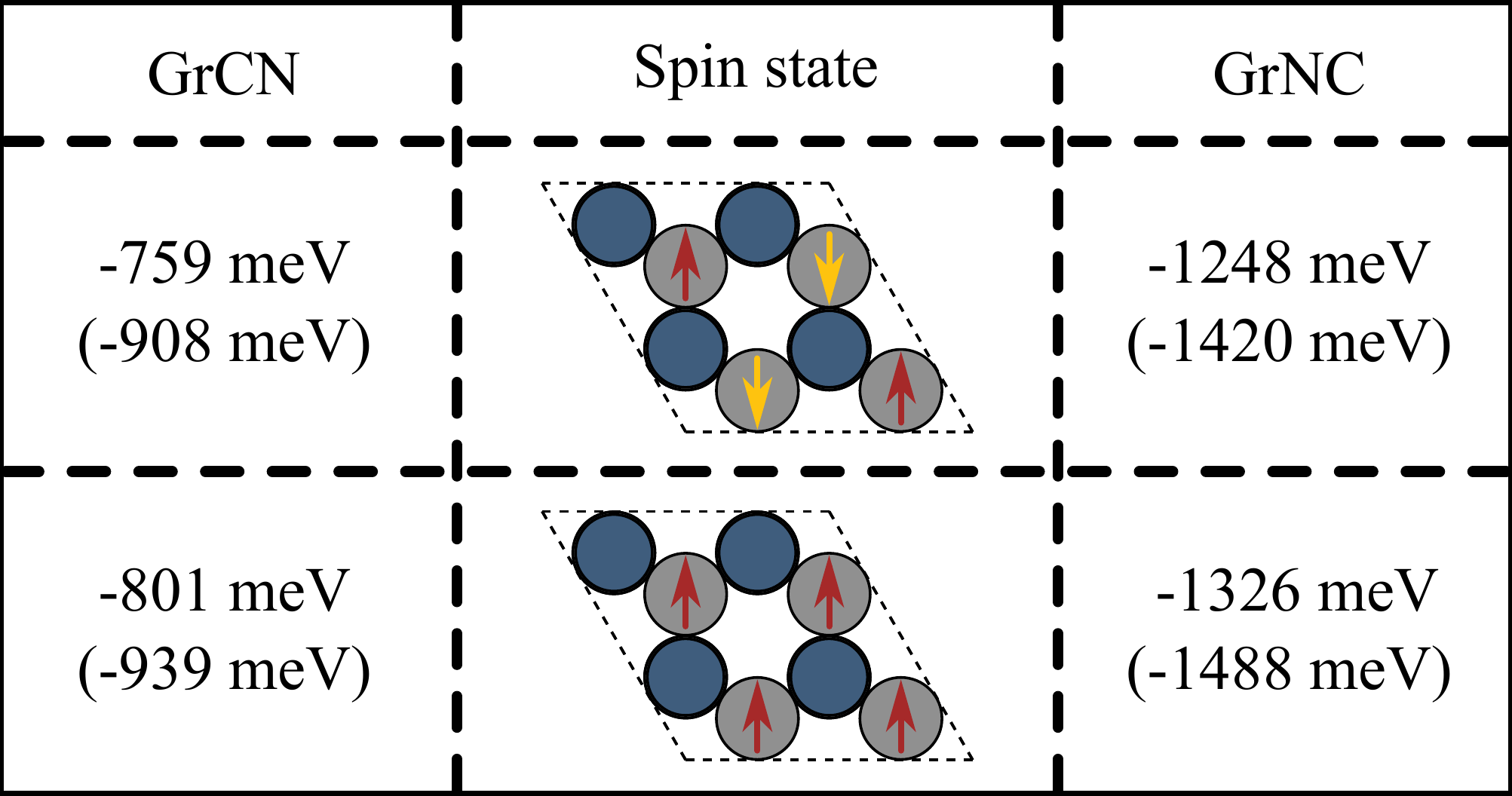}
  \caption{Stability of different spin-phases of GrCN and GrNC. The central image shows a schematic representation of the ferromagnetic as well as antiferromagnetic spin orientation. Energies are given relative to the spin-paired case with values corresponding to results obtained using the vdW-DF-CX (PBE) functional. Note that for vdW-DF-CX, the total density was used to evaluate the non-local vdW correlation.}
\label{magnetic}
\end{figure}

Given their structure, one would naturally expect these systems to display some form of long-range magnetic ordering. To confirm this, we performed vdW-DF-CX and PBE \cite{Perdew1996} calculations on a $2\times2$ supercell of GrCN and GrNC considering both ferromagnetic and antiferromagnetic ordering of the unpaired electrons and comparing their energy to that of the spin-paired state. 
The results of these calculations are depicted in fig. \ref{magnetic} which shows that, as is the case for graphone, both cyanographone (GrCN) and isocyanographone (GrNC) display a ferromagnetic ground state. This stands in contrast to the case of half-fluorinated graphene \cite{Peng2014, Boukhvalov2010} where ferromagnetic ordering of the spins does not correspond to the lowest energy state of the system. Of the two systems considered, GrCN is more stable than GrNC by $\approx 0.7$ eV/unit cell with vdW-DF-CX and PBE results differing only slightly.

\section{Stability}

One of the main challenges of partially-saturated graphene derivatives is their relatively low stability compared to fully-functionalized systems. To overcome this problem in the case of graphone, Hemmatiyan \emph{et al.} proposed depositing the material on hexagonal boron nitride (BN) \cite{Hemmatiyan2014} which they found to lead to a significant stabilization.

\begin{figure}[h]
\centering
  \includegraphics[width=0.35\textwidth]{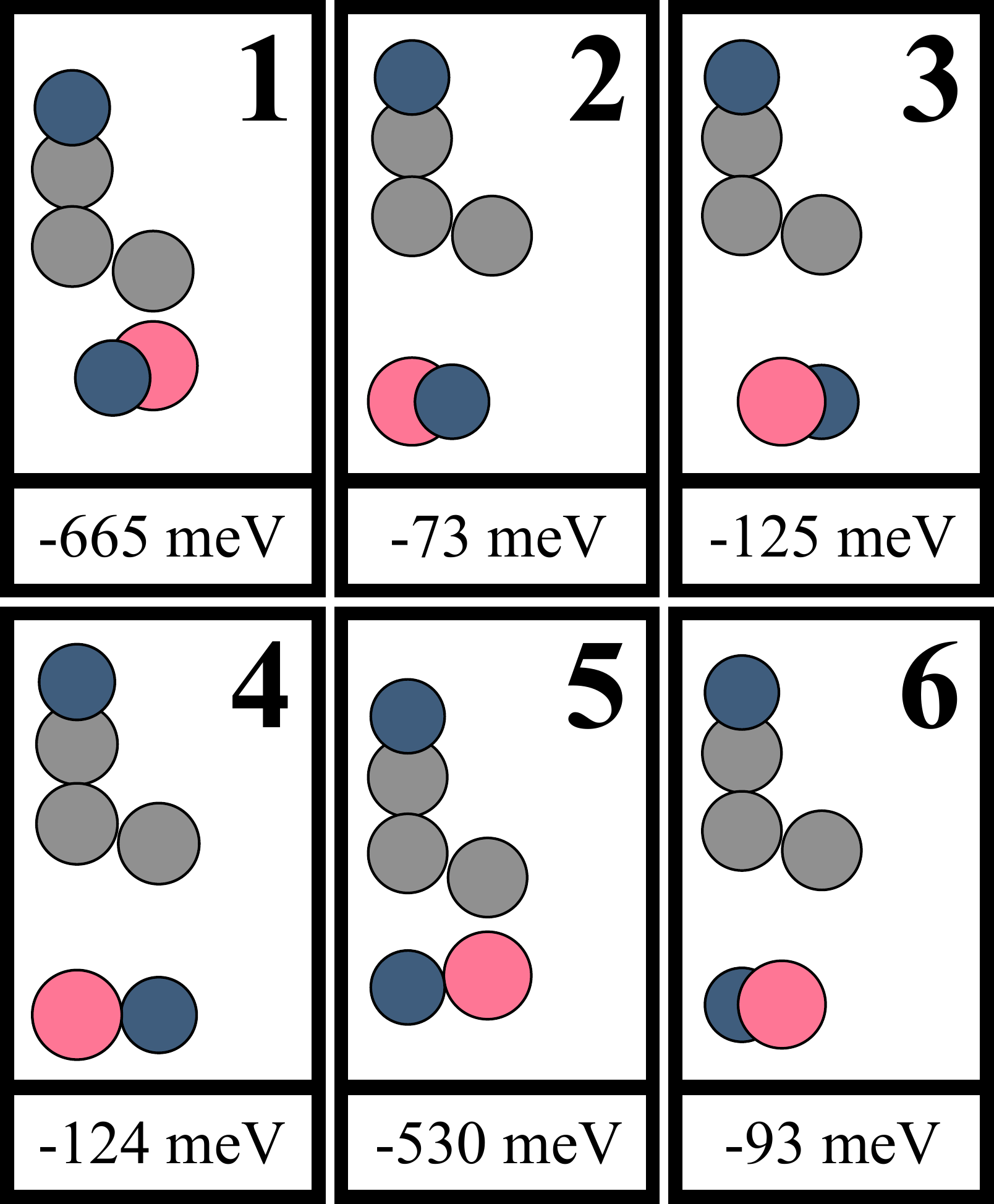}
  \caption{Relaxed structures for unit cells with different high-symmetry orientations of the GrCN@BN bilayer. Nitrogen atoms are shown in blue, while boron atoms are shown in rose and carbon atoms in gray. Note that in the case of adsorption position 1, the buckling of the graphene layer is $\approx 0.5$ $\mathring{\mathrm{A}}$ as compared to $\approx 0.4$ $\mathring{\mathrm{A}}$ in the freestanding case. Adsorption on top of a nitrogen atom has almost no influence on the buckling and e.g. structure 3 shows the same level of buckling as the freestanding GrCN layer. The binding energy for each structure is given per unit cell.}
\label{energy_orient}
\end{figure}

We mention at this point, that the lattice mismatch between GrCN/GrNC and BN is $\approx 5$\%. While the difference is rather small, it would at the same time result in long-range Moir{\'e} superlattices which might show locally different electronic properties, making practical applications of the materials more difficult. Fortunately, both cyanographone and isocyanographone bind strongly to the BN substrate (see fig. \ref{energy_orient}), making the emergence of large-scale Moir{\'e} structures less likely. In our calculations we have in all cases fixed the lattice constant of the GrCN/GrNC@BN bilayer to that of GrCN/GrNC respectively.

As a first step in considering the influence of BN adsorption on the stability of GrCN/GrNC we performed a series of calculations for different relative orientations of the two monolayers. This analysis was only done for GrCN as conclusions are expected to be applicable also to GrNC. The results -- as summarized in fig. \ref{energy_orient} -- show a clear preference for the carbon atom containing the unpaired electron to be adsorbed on top of the boron atom. The effect can be explained by the formation of a covalent bond between the electron-rich carbon atom and the relatively electron-poor boron atom within the BN layer. The position of the nitrogen atom on the other hand has a much smaller influence on the binding energy (compare orientations 1 and 5 in fig. \ref{energy_orient}). Still the system shows a clear minimum (orientation 1 in fig. \ref{energy_orient}) which is $\approx 130$ meV per unit cell more stable than the second most stable structure (orientation 5 in fig. \ref{energy_orient}). 

\begin{figure}[h]
\centering
  \includegraphics[width=0.48\textwidth]{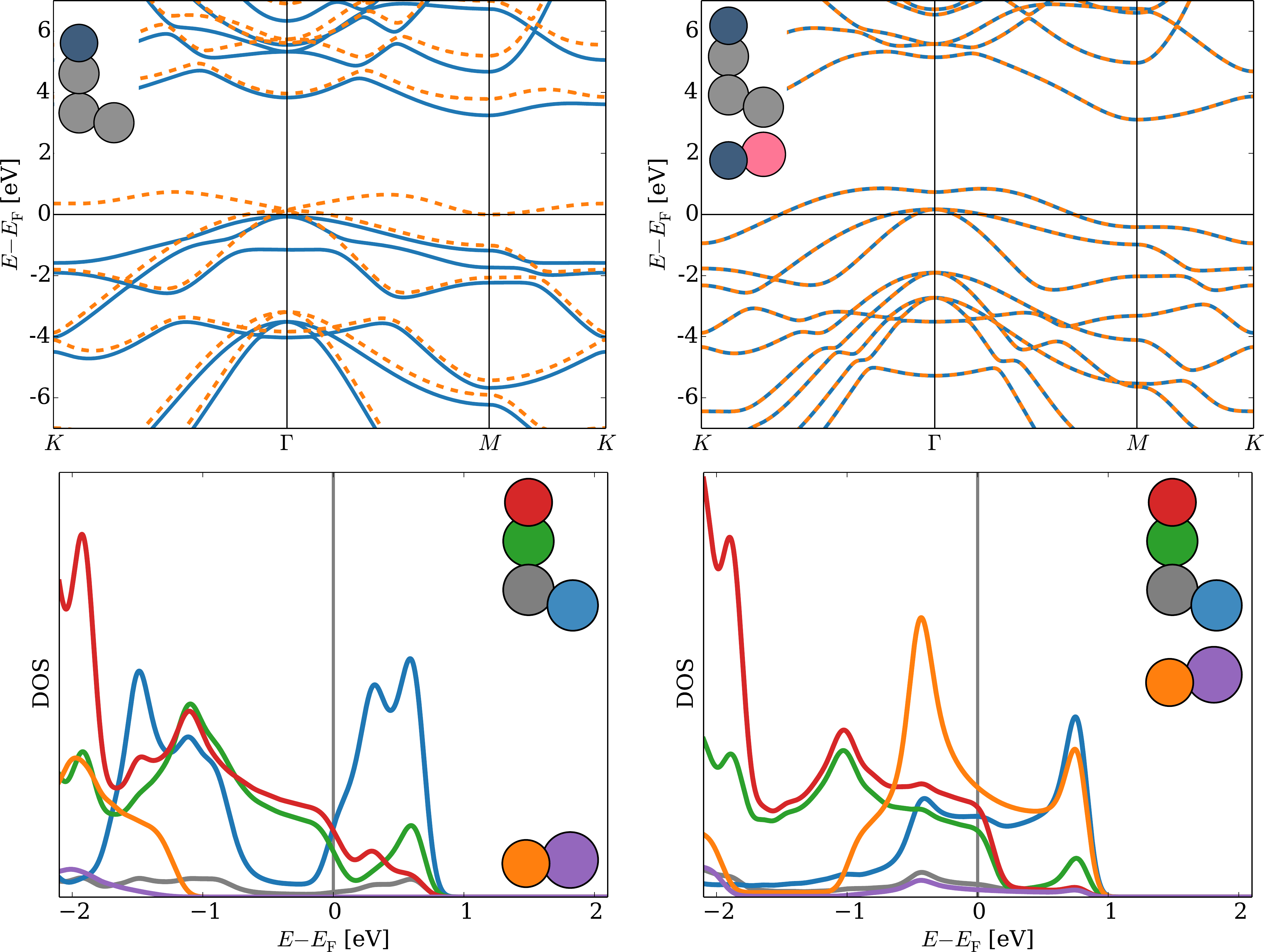}
  \caption{The two top figures show vdW-DF-CX band structures for freestanding and BN deposited cyanographone, demonstrating the loss of magnetization upon adsorption while at the same time, metallicity is conserved. Insets show the respective atomic structures and solid-blue/dashed-orange lines correspond to the majority/minority spin channels. The two bottom figures show the DOS for the same structures with curves corresponding to projections onto individual atoms. For ease of comparison between the two DOS, we have added a BN layer at a distance of 20 $\mathring{\mathrm{A}}$ from the isolated GrCN.}
\label{bands}
\end{figure}

It is worth noting, that the formation of the carbon--boron bond leads to the suppression of the GrCN magnetic moment while at the same time the system maintains its conducting nature, as is shown in fig. \ref{bands}.

Figure \ref{bands} also shows the local DOS for the two systems which provides us with some additional information as to the nature of the changes to the GrCN electronic structure upon BN adsorption. Let us focus first on the freestanding GrCN layer. Clearly, the unpaired electron on the carbon atom labeled in blue contributes most strongly to the conduction bands close to $E_{\mathrm{F}}$. The CN group (green and red curves) on the other hand, provides the largest contribution to the region immediately below the Fermi level. The BN layer, which we have added to the structure at a distance of 20 $\mathring{\mathrm{A}}$ to avoid any residual interaction between the two layers, shows no significant DOS in the vicinity of $E_{\mathrm{F}}$. 

As the two layers are put into contact, the situation changes drastically. A chemical bond is formed between GrCN and BN and the formerly unpaired electrons' contribution to the DOS around $E_{\mathrm{F}}$ decreases. At the same time the BN layer, which before did not contribute siginficantly, now shows a strong contribution on the part of the nitrogen atom (orange curve). The boron atom (violet curve) on the other hand remains close to insignificant in the valence region, contributing only very little to the overall DOS.

\begin{figure}
\centering
  \includegraphics[width=0.48\textwidth]{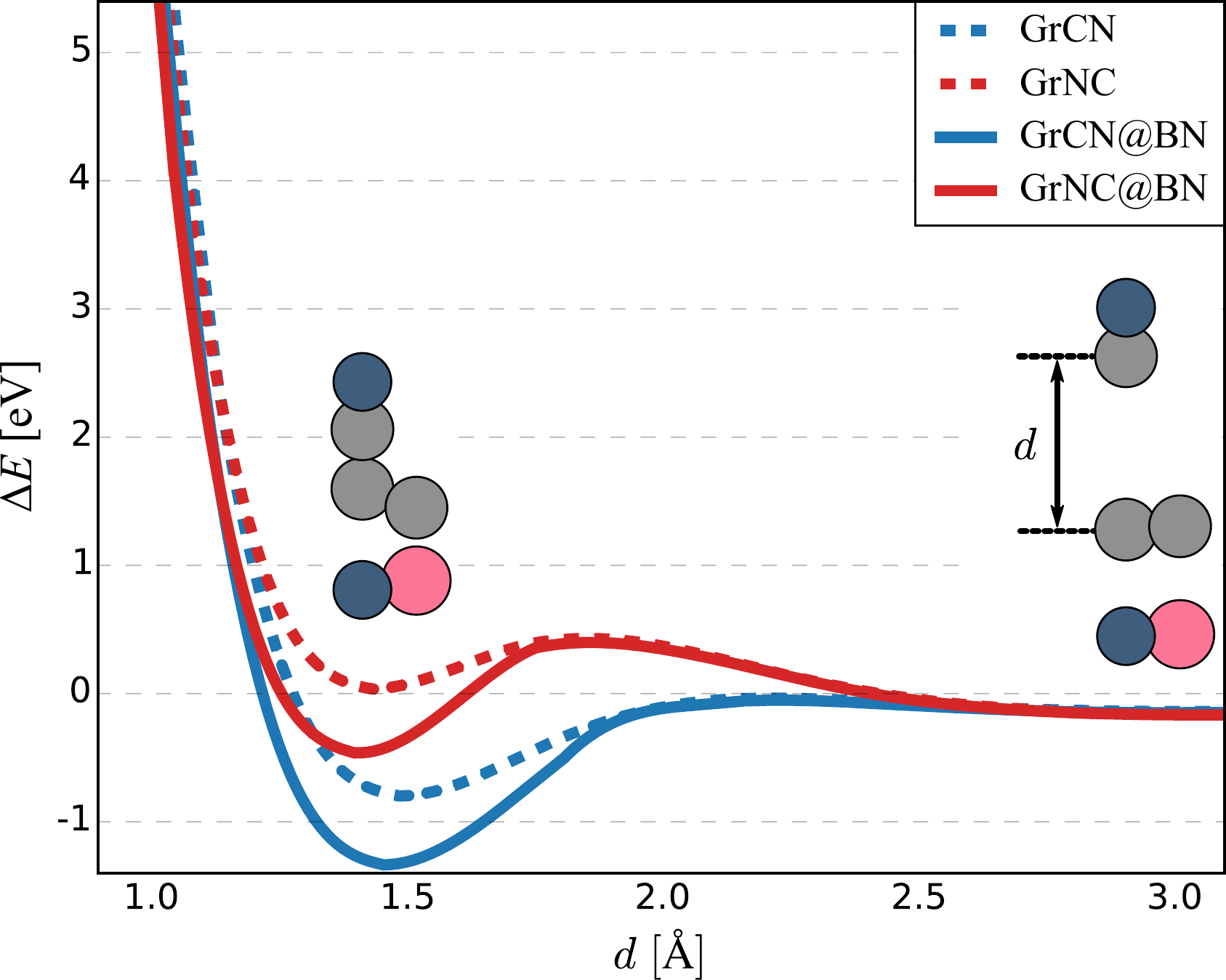}
  \caption{Potential energy surface scans for CN/NC adsorption on both freestanding (GrCN/GrNC) as well as BN adsorbed graphene (GrCN@BN/GrNC@BN). The relative orientation of graphene and BN corresponds to the most stable GrCN@BN system (orientation 1 in fig. \ref{energy_orient}). All calculations are performed using the vdW-DF-CX functional and energies are given relative to the values at $d\approx 6$ $\mathring{\mathrm{A}}$.}
\label{pes}
\end{figure}

To investigate the binding energy of CN to graphene in the case of GrCN and GrNC and to ascertain whether, as in the case of graphone, BN adsorption might be used to increase the stability of the two systems, we further performed a series of energy scans for the {\mbox{CN -- graphene}} distance. The resulting potential energy surfaces (PES) are shown in fig. \ref{pes}. We see that, as in the case of graphone, both cyanographone and isocyanographone can be strongly stabilized by adsorption on BN. While the former shows a binding energy $>1$ eV, the latter, though stabilized, still only shows very low thermodynamic stability. This being said, the energy barrier is close to 1 eV, meaning that the system could potentially show significant kinetic stability.

In the following we will nonetheless limit ourselves to considering only GrCN@BN when studying the adsorption of gas molecules since its higher thermodynamic stability makes it more likely to be stable under realistic conditions.

\section{Interactions with gas molecules}

In order to demonstrate the ability of GrCN@BN sensors to discriminate between different molecules based on whether or not they interact strongly with the materials cyano groups, we will first study the energetics of its interaction with a number of different gas molecules i.e. CO, CO$_2$, O$_2$, N$_2$, and H$_2$S. To simulate the model detector response itself, we subsequently calculate the systems current--voltage characteristics ($I-V$ curves) before and after molecular adsorption.

\begin{figure}
\centering
  \includegraphics[width=0.40\textwidth]{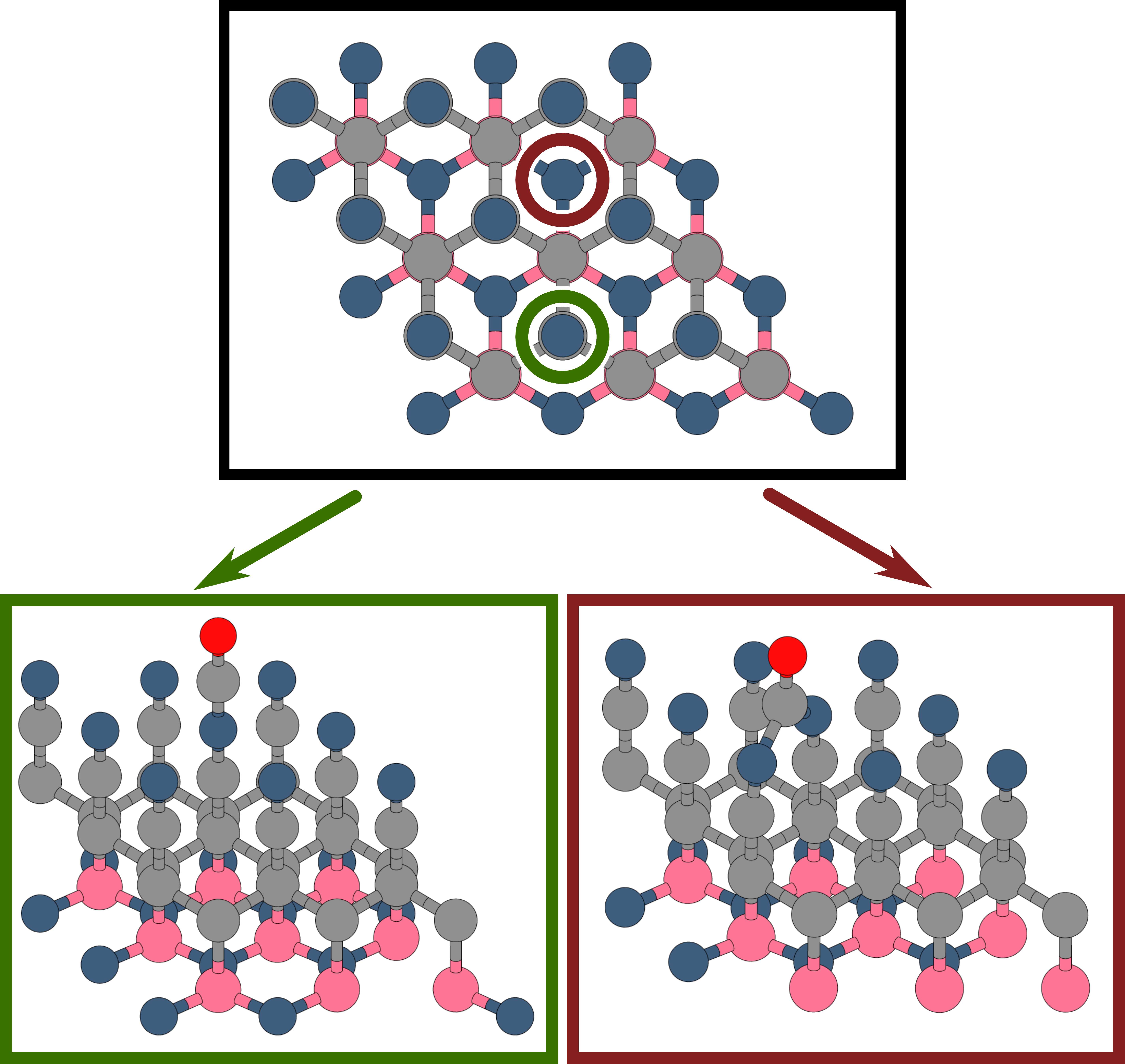}
  \caption{The black frame shows a top view of the $3\times3$ GrCN@BN supercell used in the calculations, where the two starting positions for the adsorbed molecules ($\mathbf{R}$ and $\mathbf{CN}$ in table \ref{bindign_mol}) have been indicated by red and green circles respectively. The bottom frames show the corresponding relaxed geometries for the case of CO adsorption in the P1 orientation. Note how in the case of the $\mathbf{R}$ starting position, the CO molecule has moved away from the center of the six-membered ring into a bridging position between two nitrogen atoms on neighboring CN groups.}
\label{co_adsorption}
\end{figure}

In order to study the adsorption of the different molecules on GrCN@BN, we employed a $3\times3$ supercell of GrCN@BN. For each of the five molecules studied we then considered two different initial positions, i.e., on top of a CN group as well as in the center of a six-membered ring (see fig. \ref{co_adsorption}). For each of these two initial positions we further analyzed two different orientations of the molecule. We subsequently allowed all structures to relax while keeping the unit cell parameters fixed.

\begin{table}[h]
\centering
\begin{tabular}{@{}lcccccccc@{}} 
\hline\hline
                     &&   CO   &  CO$_2$  &  O$_2$  &  N$_2$ &  H$_2$S   \\ \hline
P1$_{\mathbf{R}}$    && -1951  &  -93     &  -423   &  -31   &   -843    \\
P2$_{\mathbf{R}}$    && -83    &  -216    &  -481   &  -105  &   -817    \\\hline
P1$_{\mathbf{CN}}$   && -1283  &  -121    &  -381   &  -65   &   -805    \\
P2$_{\mathbf{CN}}$   && -151   &  -207    &  -464   &  -108  &   -2761   \\
\hline\hline
\end{tabular}
 \caption{vdW-DF-CX binding energies for CO, CO$_2$, O$_2$, N$_2$, and H$_2$S on GrCN@BN. All calculations are performed using a $3\times3$ supercell and values are given in meV. Position labels $\mathbf{R}$ and $\mathbf{CN}$ refer to starting positions of the structure optimization with the molecule either in the center of a six-membered ring ($\mathbf{R}$) or on top of a CN group ($\mathbf{CN}$), while P1/P2 have slightly different meanings in the cases of CO, H$_2$S and CO$_2$, N$_2$, O$_2$ respectively. For CO and H$_2$S, P1 and P2 indicate whether the oxygen/sulfur is pointing towards or away from the GrCN@BN surface, while for CO$_2$, N$_2$, O$_2$, P1 corresponds to the molecular axis being \emph{perpendicular} to the plane of the GrCN@BN layer while P2 refers to the axis being \emph{parallel} to it.}
\label{bindign_mol}
\end{table}

The final binding energies for all relaxed structures are shown in table \ref{bindign_mol}. We We see that both N$_2$ and CO$_2$ are only weakly adsorbed on GrCN@BN, while O$_2$, CO, and H$_2$S interact more strongly with the surface. The strongest interaction is seen for CO and H$_2$S which both form chemical bonds with the surface cyano groups, resulting in adsorption energies of $\approx 1.2 - 2.7$ eV. We will henceforth refer to this type of adsorption as \emph{chemisorption}.

\begin{figure}
\centering
  \includegraphics[width=0.40\textwidth]{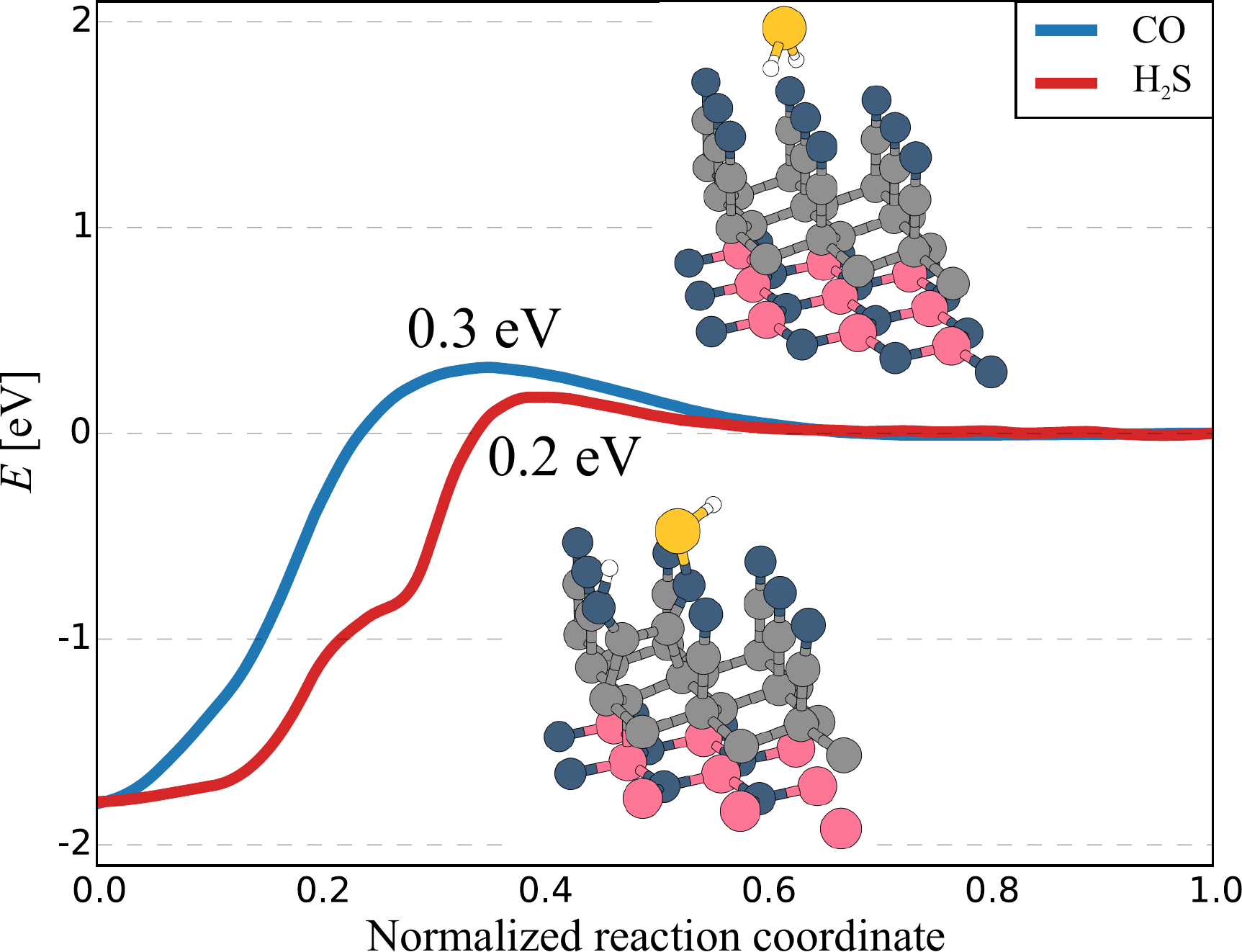}
  \caption{vdW-DF-CX potential energy curves along the reaction paths for the chemisorption of CO and H$_2$S on GrCN@BN as determined using the NEB method. Energy barriers for the individual reactions as well as representative images of the H$_2$S initial and final structures are also shown. The CO final structure corresponds to the bridged situation shown in fig. \ref{co_adsorption}. The zero-point for both curves has been set to the energy of the physisorbed minimum.}
\label{neb_pes}
\end{figure}

As already seen from fig. \ref{co_adsorption}, during the chemisorption process, the CO molecule forms one/two (depending on the orientation) chemical bonds with the surface cyano groups. For H$_2$S on the other hand, chemisorption leads to the formation of two new (N--H and N--S) bonds (see insets in fig \ref{neb_pes}). While the chemisorbed structures of both CO and H$_2$S are clearly favored energetically over their physisorbed counterparts, their relevance to the performance of a real-world detector will depend on the barriers involved in the chemisorption reaction, as high barriers might inhibit molecules from reaching the energetically favorable minimum at typical detector operating temperatures. In order to evaluate the reaction barriers for both CO and H$_2$S chemisorption on GrCN@BN, we performed NEB calculations on both molecules, starting from their lowest-energy physisorbed structures.

The results of these calculations are depicted in fig. \ref{neb_pes}. It shows that only small reaction barriers are present for the chemisorption of both CO and H$_2$S (0.3 eV and 0.2 eV respectively). While these low barriers are highly desirable in order to obtaining a good sensitivity, the large reaction energies for the two chemisorption reactions mean that desorption of the molecules is unlikely under normal operating conditions. Poisoning of the detector by CO and H$_2$S might therefore well be problematic in any real-world GrCN@BN sensor.

\section{Current--voltage characteristics}

\begin{figure}
\centering
  \includegraphics[width=0.48\textwidth]{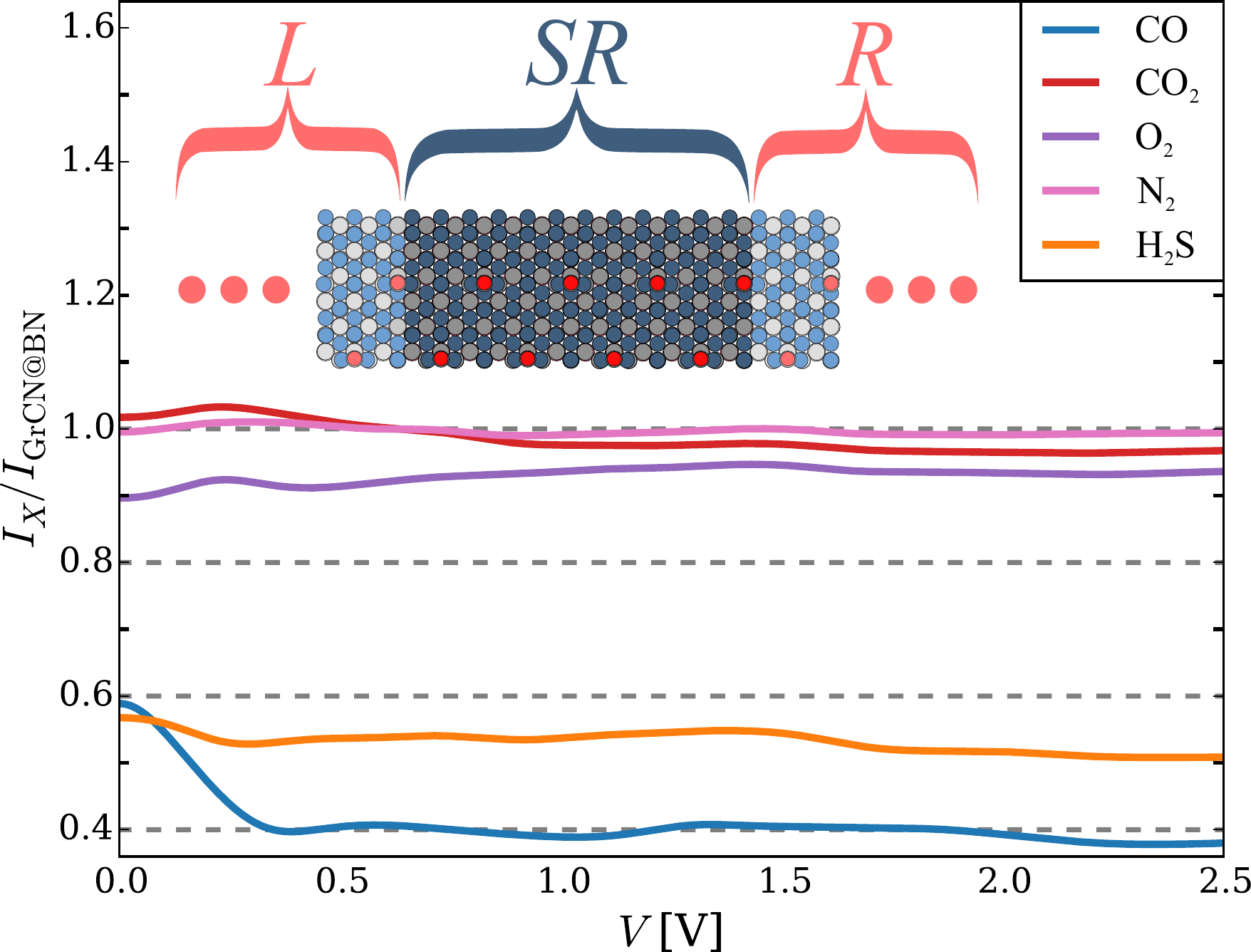}
  \caption{Voltage-dependent current-ratios ($I_{\mathrm{X}}/I_{\mathrm{GrCN@BN}}$) at $T=300$ K, where $X=$ CO, CO$_2$, O$_2$, and H$_2$S as obtained from the Landauer equation (eq. \ref{landauer}). In all cases, results refer to the most stable adsorption position of the molecule on the surface (see table \ref{bindign_mol}). The inset depicts a schematic representation of the computational setup used in the transport calculations, where the central scattering region ($\mathbf{SR}$) consists of four orthorhombic cells corresponding to the $3\times3$ supercell of (molecularly doped) GrCN@BN, depicted in fig. \ref{co_adsorption}. The desaturated regions to the left and right of the scattering region correspond to the two semi-infinite leads ($\mathbf{L}/\mathbf{R}$) and consist of a single orthorhombic cell of the same type as those constituting the scattering region.}
\label{transport}
\end{figure}

Let us now conclude our study by analyzing the simplified model of a GrCN@BN based gas sensor shown in fig. \ref{transport}. To study the ability of GrCN@BN to selectively detect adsorbed molecules in the presence of other atmospheric gases we calculate the current--voltage characteristics ($I-V$ curves) of both pristine and molecularly doped GrCN@BN. After computing the transmission using the nonequilibrium Green's function formalism, the current ($I$) can be obtained as a function of the applied voltage ($V$) from the standard Landauer equation as

\begin{equation}
I(V) = \frac{2\mathrm{e}}{\mathrm{h}} \int_{-\infty}^{\infty} \mathrm{d} E \left[ f_{\mathbf{L}}(E,\mu_{\mathbf{L}}(V)) - f_{\mathbf{R}}(E,\mu_{\mathbf{R}}(V)) \right] T(E)
\label{landauer}
\end{equation}

where $f_{\mathbf{L}/\mathbf{R}}$ are the Fermi--Dirac distributions for the left and right lead respectively and the lead chemical potentials are chosen to be $\mu_{\mathbf{L}/\mathbf{R}}=E_{\mathrm{F}}\pm V/2$ with $E_{\mathrm{F}}$ being the Fermi level. While the voltage is applied from left to right, the entire structure is periodic in the orthogonal direction which is sampled by $\mathbf{k}$-points \cite{Thygesen2005b}.

To determine whether GrCN@BN devices might be useful as chemical sensors, we are not primarily interested in the \emph{absolute} value of the current in the molecularly doped layers, but rather its \emph{relative} magnitude as compared to the pristine system ($I_{\mathrm{GrCN@BN}}$). The corresponding curves, which are shown in fig. \ref{transport}, broadly fall into three distinct groups:

\begin{enumerate}
\item Weakly bound molecules (CO$_2$ and N$_2$) largely leave the systems $I-V$ curves unaltered.
\item Moderately bound O$_2$ leads to a reduction in current of $\approx 10$\%\ at low voltages with the current more closely approaching that of weakly bound species at higher $V$.
\item Strongly bonded molecules (CO and H$_2$S) lead to significant changes in the current curves which further strongly depend on the identity of the molecule as the signals of CO and H$_2$S are well-separated ($\approx 10$ \% ) as well as largely voltage-independent at voltages $\ge 0.4$ V.
\end{enumerate}

Given these results we can draw a number of conclusions regarding the potential performance of GrCN@BN  based sensors:

\begin{enumerate}
\item Moderately bonded species such as O$_2$ could be detectable with reasonable accuracy at low $V$ ($\approx 10$\%\  reduction in the current).
\item Detection of moderately bonded O$_2$ at higher voltages ($>1$ V) becomes increasingly challenging as the current more closely approaches values observed for weakly adsorbed molecules.
\item Covalently bonded species such as CO and H$_2$S should be easily detectable ($\approx 40-60$\%\ reduction in the current) and well distinguishable ($\approx 10$\%\ separation between curves) over a range of different operating voltages though poisoning of the detector could be a problem.
\end{enumerate}

\section{Conclusions}

We have demonstrated how half-cyanation of graphene leads to a thermodynamically stable material which is conducting and shows long-range ferromagnetic ordering. This material was termed \emph{cyanographone} in analogy to half-hydrogenated graphene (graphone). We subsequently showed how the material can be further stabilized by depositing it on hexagonal BN due to the formation of chemical bonds between the unsaturated carbon sub-lattice of the cyanographone and the boron atoms within the BN layer. This adsorption further causes suppression of the ferromagnetic ordering observed for the freestanding system, without the opening of a band gap.

Lastly, we investigated the chemical sensing capabilities of this new-found material by studying the changes in its current--voltage characteristics upon adsorption of different common gas molecules. We found the system to display a highly selective sensor response to both CO and H$_2$S adsorption, as both species form a chemical bond with the GrCN@BN surface. Sensitivity to O$_2$ on the other hand is less pronounced and only present at low voltages. The application of GrCN@BN to gas detection is further made attractive by the fact that common atmospheric gases, namely N$_2$ and CO$_2$, were found to not significantly affect GrCN@BN $I-V$ curves meaning that they would not interfere with the operation of a potential GrCN@BN based sensor.

As the application of GrCN@BN to gas sensing likely by no means exhausts the entirety of its possible uses, we hope that our study, in combination with the recent advances in cyanographene synthesis, will spur more research into pseudohalogenated graphenes in general as well as cyanographone in particular, helping to fully uncover the potential of this fascinating class of materials.

\begin{acknowledgments}
We acknowledge support by the Studienstiftung des deutschen Volkes e.V., the Elsa-Neumann foundation of the Land Berlin, the Deutsche Forschungsgemeinschaft (project TR1109/2-1 and Priority Program (SPP) 1459), as well as the international Max Planck Research School "Functional Interfaces in Physics and Chemistry". The North-German Supercomputing Alliance (HLRN) and computer facilities of the Freie {\mbox{Universit\"{a}t}} Berlin (ZEDAT) are acknowledged for computer time. The authors thank Ask Hjorth Larsen (Donostia-San Sebasti{\'a}n) for helpful discussions regarding vdW-DF-CX as well as Carmen Reden (Berlin) for proofreading this manuscript.

The {\textsc{XCrySDen}} package \cite{kokalj2003computer, kokalj1999xcrysden} and \textsc{ASE} \cite{Bahn2002, LarsenASE} were used to create images of atomic structures throughout this work while plots were created using Matplotlib \cite{Hunter2007}.
\end{acknowledgments}

\bibliography{bibliography}
\end{document}